\documentclass[%
 superscriptaddress,
reprint,
showpacs,preprintnumbers,
 amsmath,amssymb,
pra,
]{revtex4-1}

\usepackage{graphicx}% Include figure files
\usepackage{dcolumn}% Align table columns on decimal point
\usepackage{bm}% bold math
\usepackage{hyperref}% add hypertext capabilities
\begin{document}

\preprint{APS/123-QED}

%\title{Quantum and Classical Perspectives on Laser-Dressed Intra-Molecular Scattering}% Force line breaks with \\
\title{Laser Dressed Scattering of an Attosecond Electron Wave Packet}% Force line breaks
%\thanks{A footnote to the article title}%

\author{Justin Gagnon}
\email{justin.gagnon@mpq.mpg.de}
\affiliation{Max-Planck-Institut f\"{u}r Quantenoptik, Hans-Kopfermann-Str. 1, D-85748 Garching, Germany}
\affiliation{Ludwig-Maximilians-Universit\"{a}t M\"{u}nchen, Am Coulombwall 1, D-85748 Garching, Germany}
\author{Ferenc Krausz}
%\email{ferenc.krausz@mpq.mpg.de}
\affiliation{Max-Planck-Institut f\"{u}r Quantenoptik, Hans-Kopfermann-Str. 1, D-85748 Garching, Germany}
\affiliation{Ludwig-Maximilians-Universit\"{a}t M\"{u}nchen, Am Coulombwall 1, D-85748 Garching, Germany}
\author{Vladislav S. Yakovlev}
\affiliation{Max-Planck-Institut f\"{u}r Quantenoptik, Hans-Kopfermann-Str. 1, D-85748 Garching, Germany}
\affiliation{Ludwig-Maximilians-Universit\"{a}t M\"{u}nchen, Am Coulombwall 1, D-85748 Garching, Germany}
\date{\today}% It is always \today, today,
             %  but any date may be explicitly specified

\begin{abstract}
%We theoretically investigate laser-dressed intra-molecular scattering -- the ejection, and subsequent scattering of an electron within a molecule, launched by an attosecond pulse under the influence of an infrared laser field. This process produces a modulated electron momentum spectrum that is understood as a result of quantum interference between two trajectories. We show that the dressing laser field can be used to control the re-scattering dynamics of an electron as it travels within the molecule, which can be observed as a change in the spectral interference pattern. However, we find that the Coulomb-Volkov approximation, a standard expression used to describe laser-dressed photoionization, cannot properly describe this process. We introduce a classical model which can quantitatively explain the laser-dressed electron spectra, notably the laser-induced changes in the spectral interference pattern.
We theoretically investigate the scattering of an attosecond electron wave packet launched by an attosecond pulse under the influence of an infrared laser field. As the electron scatters inside a spatially extended system, the dressing laser field controls its motion. We show that this interaction, which lasts just a few hundreds of attoseconds, clearly manifests itself in the spectral interference pattern between different quantum pathways taken by the outgoing electron. We find that the Coulomb-Volkov approximation, a standard expression used to describe laser-dressed photoionization, cannot properly describe this interference pattern. We introduce a quasi-classical model, based on electron trajectories, which quantitatively explains the laser-dressed photoelectron spectra, notably the laser-induced changes in the spectral interference pattern.

\end{abstract}

% PACS, the Physics and Astronomy
                             % Classification Scheme.
%Use showkeys class option if keyword
                              %display desired
\maketitle

When an electron scatters from an atom in the presence of a laser field, exotic effects can be observed which are not accessible in ordinary electron-atom scattering \cite{Mittleman:1982}. Laser-assisted electron-atom collisions were mostly studied using monochromatic electron and laser beams \cite{Ehlotzky:1998}. In such experiments, an atom and an electron take part in a collision at some random time in a monochromatic laser field, which acts as a perturbation to the scattering event. Recent progress in the generation of few-cycle laser pulses \cite{Baltuska:2003} has enabled a new class of experiments, where an electron wave packet is \emph{coherently} launched, e.g. by strong-field ionization or by an attosecond light pulse, and steered in the field of an intense laser pulse \cite{Corkum:1989,Itatani:2002,Kienberger:2004}. This remarkable degree of control over electron motion, together with the attosecond timing precision between the light field and the electron wave packet have enabled fundamental experimental \cite{Drescher:2002,Niikura:2002,Itatani:2004,Baker:2006} and theoretical \cite{Lucchese:1982,Lein:2002a,Yudin:2007,Grafe:2008} developments in atomic, molecular, and optical physics.
%Having been launched over a duration much shorter than the laser period, the electron wave packet can be steered by the laser field oscillations.

In this paper, we theoretically study how a laser field affects the scattering of an attosecond electron wave packet as it travels inside a spatially extended system. Experimentally, this can be realized by ionizing a localized electronic state of a molecule with an attosecond extreme-ultraviolet (XUV) pulse. As it exits the molecule, the photoelectron will be scattered by other atoms in the molecule before heading to the detector. In particular, the scattering of a photoelectron within a molecule has recently attracted a significant amount attention on its own: in addition to the Cohen-Fano oscillations \cite{Cohen:1966}, ionization from a localized core orbital of CO also produces a modulation in the momentum-resolved cross sections \cite{Zimmermann:2008} arising from the interference between trajectories taken by the outgoing electron, referred to as \emph{intra-molecular scattering}. The interference pattern produced at the detector can be interpreted as a holographic image \cite{Krasniqi:2010}, and can be used to retrieve the molecular structure seen by the outgoing electron. In this paper, we show that a near-infrared (NIR) laser wave form, temporally synchronized to the collision event, can be used to control the paths taken by the outgoing photoelectron, which can be observed by measuring the interference in the photoelectron spectrum.

\begin{figure}
  % Requires \usepackage{graphicx}
  \includegraphics[width=75mm]{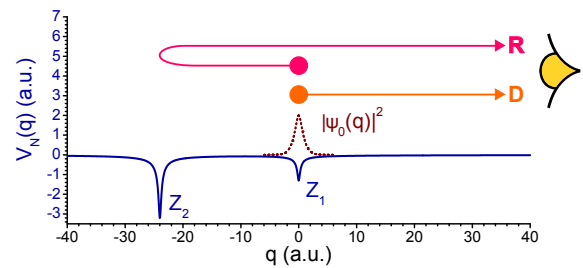}
  \caption{The ionic potential (solid line) dips at $q=0$ and $q=-24\,\textrm{a.u.}$ The initial state (dotted line) is localized at $q=0$. The reflected and direct trajectories are labeled with ``R'' and ``D''.}\label{ims}
\end{figure}
%The system we consider is an electron moving in the potential of a one-dimensional diatomic molecule with fixed nuclei, driven by the electromagnetic fields of an attosecond XUV pulse, used for ionization, and a femtosecond laser pulse, used to control the outgoing electron. We choose the initial (bound) state $|\psi_0\rangle$ to be the first excited state of our potential (ionization potential $W\approx 12.17\,\textrm{eV}$) because it emulates a core molecular shell localized on the shallower potential well (Fig. \ref{ims}), yielding a larger IMS probability. The Hamiltonian of the electron interacting with the molecular potential and the electromagnetic radiation, assuming the dipole approximation, is
Since we are interested in the most general features of laser-dressed scattering, we consider a one-dimensional model system, akin to a nanometer-scale Fabry-P\'{e}rot etalon for the free electron wave packet. Our system is composed of two potential wells chosen such that the electron is initially localized in one of them (Fig. \ref{ims}). The initial (bound) state $|\psi_0\rangle$ is the first excited state of the double-well system (ionization potential $W\approx 12.17\,\textrm{eV}$). The Hamiltonian of the electron interacting with the potentials and the electromagnetic radiation, assuming the dipole approximation, is
\begin{eqnarray}\label{hamiltonian}
% \nonumber to remove numbering (before each equation)
  H &=& \frac{p^2}{2}+V_\mathrm{N}(q)+\Big(F_\mathrm{L}(t)+F_\mathrm{X}(t)\Big)\,q,\\\nonumber
  V_\mathrm{N}(q) &=& \frac{1}{Z_1+Z_2}\left(-\frac{Z_1}{\sqrt{q^2+a^2}}-\frac{Z_2}{\sqrt{(q-q_\mathrm{r})^2+a^2}}\right)
\end{eqnarray}
with $a\approx0.2236\,\textrm{a.u.}$, $q_\mathrm{r}=-24\,\textrm{a.u.}$, $Z_1=2\,\textrm{a.u.}$ and $Z_2=5\,\textrm{a.u.}$ $p$ is the electron's momentum, $V_\mathrm{N}(q)$ is the potential due to the nuclei, with nuclear charges $Z_1$ and $Z_2$. $F_\mathrm{L}(t)$ and $F_\mathrm{X}(t)$ represent the electric fields of the NIR laser and attosecond XUV pulses, respectively. They are given explicitly by
\begin{eqnarray}
  \label{xuvfield}
  F_\mathrm{X}(t)&=&10^{-5}\int_{-\infty}^{\infty}G_{\kappa,\theta}(\omega-W)\cos{(\omega t)}\,\mathrm{d}\omega,\\
  %\textrm{with }\kappa&\approx&17.94,\textrm{ and }\theta\approx 0.1735.
  \label{laserfield}
  F_\mathrm{L}(t)&=&-\frac{\mathrm{d}}{\mathrm{d}t}\left(\frac{F_0}{\omega_\mathrm{L}}\cos^4{\left(\frac{\pi t}{2\tau_\mathrm{L}}\right)\sin{(\omega_\mathrm{L}t)}}\right),\textrm{ }|t|\leq\tau_\mathrm{L},
  %F_\mathrm{L}(t)&=&
%  \left\{
%  \begin{array}{cll}
%    -\frac{\mathrm{d}}{\mathrm{d}t}\left(\frac{F_0}{\omega_\mathrm{L}}\cos{\left(\frac{\pi t}{2\tau_\mathrm{L}}\right)\sin{(\omega_\mathrm{L}t)}}\right)&,&|t|<\tau_\mathrm{L}\\
%    0&,&\textrm{ otherwise}
%  \end{array}
%  \right.
\end{eqnarray}
while $F_\mathrm{L}=0$ for $|t|>\tau_\mathrm{L}$. The XUV spectrum is $G_{\kappa,\theta}(\omega-W)$, the Gamma distribution with mean $\kappa\theta$ and variance $\kappa\theta^2$. The Gamma distribution was chosen to avoid populating Rydberg states. The parameters $\kappa$ and $\theta$ produce an XUV spectrum peaked at an energy $\omega_\mathrm{X}=80\,\textrm{eV}+W=92.2\,\textrm{eV}$, with a FWHM bandwidth $\delta\omega_\mathrm{X}\approx 32.4\,\textrm{eV}$, yielding a $55.4\,\textrm{as}$ pulse. The attosecond XUV pulse temporally overlaps with the center of the laser pulse, at the extremum of $F_\mathrm{L}(t)$. $\tau_\mathrm{L}$ and $\omega_\mathrm{L}$ are chosen to produce a laser pulse with a full-width at half-maximum (FWHM) duration of $3\,\textrm{fs}$ and a central wavelength of $800\,\textrm{nm}$; $F_0$ is the laser field's peak amplitude. The laser electric field $F_\mathrm{L}(t)$, which we refer as the \emph{control field}, is therefore a cosine pulse. The laser field amplitude $F_0$ is the variable parameter in our analysis. It influences the trajectory taken by the electron inside the system.

In the absence of a laser field, $F_0=0$, the solution of the time-dependent Schr\"{o}dinger equation (TDSE) can be formally written for positive energies, and up to a constant phase, as
\begin{eqnarray}
    \label{smatrix}
    \langle p|\psi(t_\mathrm{f})\rangle&=&\mathrm{e}^{-\frac{\mathrm{i}}{2}p^2 t_\mathrm{f}}\int_{-\infty}^{t_\mathrm{f}}F_\mathrm{X}(t)d(p)e^{\mathrm{i}\big(\frac{p^2}{2}+W\big)t}\mathrm{d}t,\\
    \label{freefree}
    \textrm{assuming }&&F_\mathrm{X}(t)\langle\phi_{p_2}| q|\phi_{p_1}\rangle\approx0,
\end{eqnarray}
and $W$ is the ionization energy. The dipole matrix elements $d(p)=\langle\phi_p| q|\psi_0\rangle$ are evaluated between the initial (bound) state $|\psi_0\rangle$ and positive-energy eigenstates $\langle\phi_p|$ of the Hamiltonian $H_0=p^2/2+V_\mathrm{N}(q)$, with asymptotic momenta $p$. The eigenstates $|\phi_p\rangle$ satisfy the Lippmann-Schwinger equation with an advanced Green's function,
\begin{equation}\label{LS}
    \langle q|\phi_p\rangle=e^{\mathrm{i}p q}-\frac{2\mathrm{i}}{|p|}\int_{-\infty}^{\infty}e^{-\mathrm{i}|p(q-q')|}V_\mathrm{N}(q')\phi_p(q')\mathrm{d}q',
\end{equation}
which we solve by computing the Born series until convergence.

Now, for the system under present scrutiny (Fig. \ref{ims}), laser-dressed scattering is clearly more pronounced on the right-going wave packet. It manifests itself as a modulation of the photoelectron spectrum for positive momenta, corresponding to a characteristic length of $\approx 49.3\,\textrm{a.u.}$, which is about twice the internuclear spacing. Quantum-mechanically, this modulation is explained by the fact that the matrix elements $d(p)$ are spectrally modulated for positive momenta.
%\begin{figure}
%  % Requires \usepackage{graphicx}
%  \includegraphics[width=80mm]{f2.eps}
%  \caption{The electron energy spectrum in the left and right channels (solid lines) result from the response of the system, quantified by the dipole transition matrix elements (dash-dotted lines), to the driving XUV field. The shaded areas represent the XUV spectra (for negative and positive frequencies), shifted by the ionization potential $W$.}\label{dme}
%\end{figure}
In the presence of a NIR laser field, there is a standard amendment to (\ref{smatrix}), which can describe laser-dressed single-photon ionization. It is the Coulomb-Volkov approximation (CVA) \cite{Duchateau:2002,Kornev:2002}. The CVA is used to account for the action of the laser field on the ejected electron. We use the version of CVA that reads
\begin{eqnarray}
    \label{cva}
    \langle p|\psi_\mathrm{CVA}(t_\mathrm{f})\rangle&=&\mathrm{e}^{-\frac{\mathrm{i}}{2}p^2 t_\mathrm{f}}\int_{-\infty}^{t_\mathrm{f}}F_\mathrm{X}(t)d(p+A(t))\nonumber\\
    &\times&\mathrm{e}^{\mathrm{i}\left(-\frac{1}{2}\int_t^{t_\mathrm{f}}\left(p+A(t')\right)^2\mathrm{d}t'+Wt\right)}\mathrm{d}t,
\end{eqnarray}
where $A(t)$ is the vector potential of the NIR field. The Coulomb-Volkov approximation relies on a couple of intuitive arguments. First, an electron trajectory ending with a momentum $p$, e.g. at the detector, must have been launched with an energy $\left(p+A(t)\right)^2/2$ at the moment of ionization $t$. Therefore, $\langle\phi_{p}|$ should back-propagate to $\langle\phi_{p+A(t')}|$ at the moment of ionization, which is why the matrix element $d(p+A(t))$ is used in (\ref{cva}). Second, the electron's evolution under the laser field is accounted for by the Volkov phase, which is the quantum phase acquired by a free electron in an electromagnetic field.

\begin{figure}
  % Requires \usepackage{graphicx}
  \includegraphics[width=75mm]{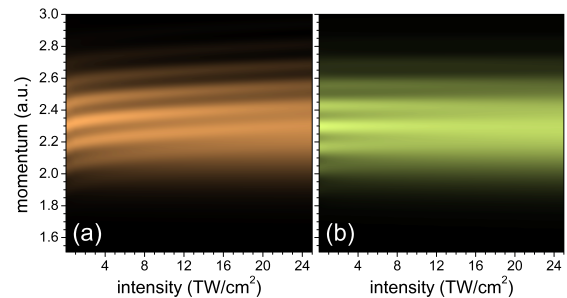}
  \caption{Laser-dressed photoelectron spectra numerically evaluated (a) from the temporal Schr\"{o}dinger equation and (b) from the Coulomb-Volkov approximation. The CVA does predict a fringe pattern, but fails to account for its change under the laser field.}\label{cvaims}
\end{figure}

The CVA approximation (\ref{cva}) is known to adequately describe laser-dressed photoionization of atoms \cite{Duchateau:2002}. However, it cannot properly describe the laser-dressed photoionization if the system is too large. In Figure \ref{cvaims}, a series of laser-dressed photoelectron spectra computed by numerically solving the TDSE (panel a) are compared to those obtained by evaluating the CVA expression (\ref{cva}), for different values of the control field's intensity (panel b). Since the attosecond XUV pulse is centered at the peak of the laser electric field, hardly any momentum shift is expected for the outgoing electron. Nevertheless, the TDSE predicts a noticeable shift of the interference pattern to larger momenta as a function of the strength $F_0$ of the control field. This effect is not accounted for by the CVA. Now, since the CVA is a semi-classical modification of the quasi-exact expression (\ref{smatrix}), this would appear to preclude an intuitive classical interpretation of laser-dressed photoelectron scattering based on the simple trajectories shown in Figure \ref{ims}.

To describe laser-dressed photoelectron scattering, we present an intuitive theoretical model based on trajectories \cite{Smirnova:2008,Heller:1981}. As will be clear from the subsequent analysis, our model quantitatively accounts for the effects of the laser field on the spectral interference. In order to gain a deeper understanding of the dynamics of the electron as it exits the system, we take a closer look at the final positive-energy components of the electron's wave function. It is useful to think of the propagated state $|\psi(t_\mathrm{f})\rangle$ as being composed of a sum of two two parts corresponding to two sets of trajectories taken by the outgoing electron: those for which the electron heads directly to the detector, and those which make it re-scatter off the adjacent nucleus before going to the detector. For our subsequent analysis, we further simplify this picture into a rather classical one by considering strictly two trajectories: a \emph{direct} trajectory and a \emph{reflected} trajectory, which will be described below.

We can separate these two trajectories from the total positive-energy wave packet by applying a simple unitary transformation corresponding to the back-propagation of a free particle:
\begin{equation}
    \label{unitary_transformation_wave_parcel}
    |w(t_\mathrm{f})\rangle=\exp\left(\frac{\mathrm{i}}{2}p^2 t_\mathrm{f}\right)|\psi(t_\mathrm{f})\rangle.
\end{equation}
%\begin{equation}
%    \label{unitary_transformation}
%    |w\rangle=\exp\left(\frac{\mathrm{i}}{2}p^2 t_\mathrm{f}\right)|\psi(t_\mathrm{f})\rangle.
%\end{equation}
\begin{figure}
  % Requires \usepackage{graphicx}
  \includegraphics[width=75mm]{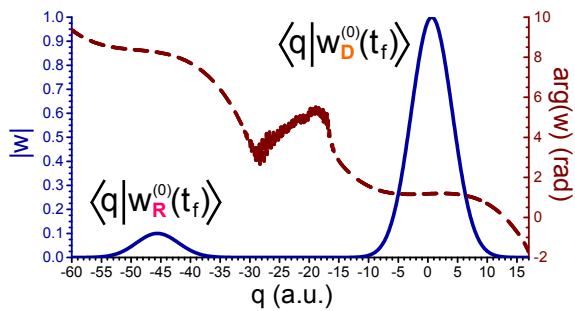}
  \caption{The amplitude (solid line) and phase (dashed line) of the wave parcel shows two separate contributions $|w^{(0)}_\mathrm{R}(t_\mathrm{f})\rangle$ and $|w^{(0)}_\mathrm{D}(t_\mathrm{f})\rangle$. The wave parcel is evaluated in the absence of the control field, as indicated by the superscript ``$(0)$''. We removed the central momentum from the wave parcel to better visualize the phases of the direct and reflected components. Once projected into real space, the wave parcels $w^{(0)}_\mathrm{D}(t_\mathrm{f},q)$ and $w^{(0)}_\mathrm{R}(t_\mathrm{f},q)$ are located at the \emph{apparent} starting points of the electron trajectories.}\label{wparcel_fig}
\end{figure}
The projection of $|w(t_\mathrm{f})\rangle$ in configuration space allows us to define a quantity which we call the \emph{wave parcel},
\begin{eqnarray}\label{wparcel}
    w(t_\mathrm{f},q)&=&\int_{-\infty}^{\infty}\langle p|\psi(t_\mathrm{f})\rangle e^{\frac{\mathrm{i}}{2}p^2 t_\mathrm{f}}e^{\mathrm{i}p q}\mathrm{d}p;
    %\textrm{and }\tilde{w}(p)&=&e^{\mathrm{i}\frac{p^2}{2}t}\langle p|\psi(t)\rangle.
\end{eqnarray}
the states $\langle p|$ are just free-particle eigenstates. The field-free wave parcel $w^{(0)}(t_\mathrm{f},q)$ is plotted in Fig. \ref{wparcel_fig}. It has been evaluated with the control field turned off ($F_0=0$) as indicated by the superscript ``$(0)$''. The wave parcel clearly consists of two parts. It has a large hump, denoted by $w^{(0)}_\mathrm{D}(t_\mathrm{f},q)$ in Fig. \ref{wparcel_fig}, centered about the origin, and a smaller hump $w^{(0)}_\mathrm{R}(t_\mathrm{f},q)$ centered at $\sim-45.6\,\textrm{a.u.}$ As will be clear from the subsequent analysis, the large and small humps correspond, respectively, to the direct and reflected trajectories taken by the outgoing electron.

Since the wave parcel is obtained by propagating the final positive-energy wave function backward in time as a free particle, the position of the wave parcel represents the \emph{apparent} starting point of the electron from the perspective of an observer measuring the electron at the final time $t_\mathrm{f}$. This apparent starting position is analogous to a \emph{delay} of the wave parcel. Undoing the unitary transformation to the individual wave parcel components thus gives the direct and reflected wave packets at the end of the propagation, i.e. $|\psi_\mathrm{R,D}(t_\mathrm{f})\rangle=\exp\left(-\frac{\mathrm{i}}{2}p^2 t_\mathrm{f}\right)|w_\mathrm{R,D}(t_\mathrm{f})\rangle$.

Here we also note an important property of the wave parcel. For $t_\mathrm{f}$ sufficiently large, the positive-energy part of the wave function essentially propagates as a free particle, rendering the associated wave parcel time-independent,
\begin{equation}
  \label{wave_parcel_limit}
    \lim_{t_\mathrm{f}\rightarrow\infty}\frac{\mathrm{d}}{\mathrm{d}t}|w(t_\mathrm{f})\rangle=0.
\end{equation}
Thus, we henceforth drop the time argument of the wave parcel because we assume the electron is measured long after its interaction with the fields and the ionic potential, so that its wave parcel $|w\rangle$ no longer depends on the time of measurement.

%The final coordinates (position and velocity) we extract from these segregated field-free wave packets are then used for calculating the classical trajectories.
\begin{figure}
  % Requires \usepackage{graphicx}
  \includegraphics[width=75mm]{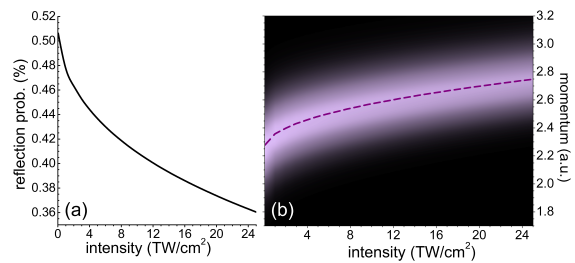}
  \caption{In panel a, the reflection probability is plotted against the control field intensity, while panel b shows the momentum spectra of the reflected wave parcel, along with the classically-expected final momentum of the reflected trajectory (dashed line). }\label{reflected}
\end{figure}

For the system under consideration, the net momentum shift of the direct trajectory is negligible since the ionization takes place at a zero-crossing of the control field's vector potential. The interesting physics occurs during the re-scattering event, experienced by the smaller hump of the wave parcel (the reflected wave parcel). Figure \ref{reflected} shows the parameters of the reflected wave parcel as a function of the intensity of the control field. Since photoionization takes place under a positive laser field the rescattered electron is initially accelerated toward the adjacent potential. Since the rescattering probability decreases with larger incident momentum, the probability of reflection naturally decreases with the control field strength, as displayed in Fig. \ref{reflected}-a. Furthermore, the spectra of the reflected wave parcel, shown in Fig. \ref{reflected}-b, are progressively shifted to larger momenta for increasing strengths of the control field.

%To this end, we introduce our classical interpretation of IMS. Just as the CVA explains laser-dressed spectra by amending the field-free expression (\ref{smatrix}) to account for the action of the laser field, we explain the laser-dressed spectra by adjusting the field-free wave packet using classical arguments. We consider the propagation of two trajectories: one heading directly to the detector (the \emph{direct} trajectory), and another which non-classically reflects off the adjacent nucleus before going to the detector (the \emph{reflected} trajectory). First, we define a quantity which we call the \emph{wave parcel}:

%The wave parcel we obtain from (\ref{smatrix}), for the field-free case $F_0=0$, consists of two blobs. Figure \ref{wparcel} shows wave parcels for negative, positive and zero dressing laser fields. As will be demonstrated from the subsequent analysis, the large and small blobs correspond respectively to the direct and reflected trajectories. This assertion is supported by the fact that the reflected component of the wave parcel for a negative (positive) dressing field is larger (smaller) compared to the field-free reflected part. This is because the negative laser field decelerates the IMS wave packet, whereas the positive field accelerates the IMS wave packet as it travels within the molecule before hitting the adjacent potential well, for which the reflection probability decreases with incident momentum.

The dependence of the reflected momentum on the laser field is well explained by classical mechanics. The dashed line plotted in Fig. \ref{reflected}-b represents final momenta computed by classically propagating an electron along the reflected trajectory. We conduct calculations of both the reflected (R) and direct (D) trajectories by launching them at the center of the initial potential ($Z_1$), at $x_\mathrm{R,D}(0)=0$, with initial velocities
\begin{equation}
    \label{intitial_velocity_reflected}
    v_j(0)=\pm\sqrt{\left(v^{(0)}_j(t_\mathrm{f})\right)^2+2\Big(V_\mathrm{N}\big(x^{(0)}_j(t_\mathrm{f})\big)-V_\mathrm{N}(0)\Big)},
\end{equation}
where the subscript $j$ stands for the direct (D) or reflected (R) trajectory, and the final positions $x^{(0)}_j(t_\mathrm{f})=\langle\psi^{(0)}_j(t_\mathrm{f})|q|\psi^{(0)}_j(t_\mathrm{f})\rangle$ and velocities $v^{(0)}_j(t_\mathrm{f})=\langle\psi^{(0)}_j(t_\mathrm{f})|p|\psi^{(0)}_j(t_\mathrm{f})\rangle$ are extracted from the \emph{field-free} reflected wave packets $|\psi^{(0)}_j(t_\mathrm{f})\rangle$. The $\pm$ sign in (\ref{intitial_velocity_reflected}) indicates that direct trajectories are launched to the right ($+$) while reflected trajectories are launched to the left ($-$). The re-scattered electron thus initially travels to the left toward the adjacent potential ($Z_2$). Once inside the scattering potential, at $x_\mathrm{R}=-24\,\textrm{a.u.}$, the electron abruptly reverses its direction as if it elastically bounces off a wall, leading it back towards the detector. In the absence of the control field, the reflected electron crosses the initial potential $Z_1$ $\sim 480\,\textrm{as}$ later. When the control field is turned on, this delay changes by $\pm 20\,\textrm{as}$ for the field intensities considered herein.

We can also explain the change in the fringe pattern as a function of the control field strength. Just as the CVA explains laser-dressed spectra by amending the field-free expression (\ref{smatrix}) to account for the action of the laser field, we explain the laser-dressed spectra by adjusting the field-free wave parcel using our classically evaluated trajectories.

To explain the laser-dressed interference pattern, we need to consider both the direct and reflected trajectories. We launch the direct trajectories of the classical particle at the center of the initial potential ($Z_1$), at position $x=0$ with initial velocity given by (\ref{intitial_velocity_reflected}). The classical simulations produce the analogues of the phases $\Delta S_{\mathrm{R},\mathrm{D}}^{(\mathrm{L})}$, momenta $p_{\mathrm{R},\mathrm{D}}^{(\mathrm{L})}$ and positions $q_{\mathrm{R},\mathrm{D}}^{(\mathrm{L})}$ of the direct (D) and reflected (R) wave parcels. These three quantities set, respectively, the position, the bias and the spacing of the interference pattern in the photoelectron spectrum, and are deduced from the classical trajectories according to the following relations:
\begin{eqnarray}
% \nonumber to remove numbering (before each equation)
\label{wp_pos_mom}
q_j^{\mathrm{(L)}}&=&x_j(t_\mathrm{f})-v_j(t_\mathrm{f})t_\mathrm{f},\textrm{ }p_j^{\mathrm{(L)}}=v_j(t_\mathrm{f}),\textrm{ }\\
\label{wp_act}
\Delta S_j^{\mathrm{(L)}}&=&\int_{0}^{t_\mathrm{f}}{L\big(x_j(t),v_j(t),t\big)\mathrm{d}t}-\frac{1}{2}v_j^2(t_\mathrm{f})t_\mathrm{f},
%\textrm{and }x&=&x_0+\int_{t_0}^{t}{\int_{t_0}^{t'}{f(t'')\mathrm{d}t''}\mathrm{d}t'}\\
%&-&\int_{t_0}^{t}{f(t')\mathrm{d}t'}(t-t_0)
\end{eqnarray}
and $L(x_j(t),v_j(t),t)$ denotes the Lagrangian evaluated along the reflected or direct trajectory, parameterized by $x_j(t)$ and $v_j(t)$. Again, the index $j\in\{\mathrm{R},\mathrm{D}\}$ refers to the direct (D) or reflected (R) trajectory. Since the wave parcel is obtained by back-propagating the wave packet as a free particle, the classical parameters $\Delta S_j^{\mathrm{(L)}}$, $p_j^{\mathrm{(L)}}$, and $q_j^{\mathrm{(L)}}$ also include the effects of free-particle back-propagation.

\begin{figure}
  % Requires \usepackage{graphicx}
  \includegraphics[width=75mm]{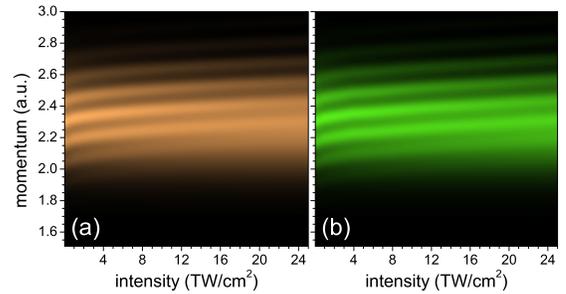}
  \caption{The classically-adjusted laser-dressed photoelectron spectra (b), based on the transformation (\ref{classical_adjustment}) reproduce the correct fringe patterns predicted by the TDSE (a). It is the difference in the back-propagated action $\Delta S$, between the reflected and direct trajectories, that sets the position of the fringes. The classically-adjusted spectra shown in panel b above are to be compared to those evaluated from the CVA (Fig. \ref{cvaims}-b).}\label{clims}
\end{figure}

Using the classical quantities, we explain the laser-dressed photoelectron spectrum by modifying the field-free direct and reflected wave parcels, $w_\mathrm{D}^{(0)}(q)$ and $w_\mathrm{R}^{(0)}(q)$ respectively. We obtain the laser dressed wave parcels $w_{\mathrm{R},\mathrm{D}}^{(\mathrm{L})}(q)$ according to the prescription
\begin{eqnarray}
  \label{classical_adjustment}
    %w_j^{(\mathrm{L})}(q)&=&\int_{-\infty}^{\infty}\tilde{w}_j^{(0)}(p)\mathrm{e}^{\mathrm{i}p\big(q-q_j^{\mathrm{(L)}}+q_j^{(0)}\big)}\mathrm{d}p\\
    %\nonumber
    w_j^{(\mathrm{L})}(q)&=&w_j^{(0)}\left(q-q_j^{(\mathrm{L})}+q_j^{(0)}\right)\\
    \nonumber
    &\times&\mathrm{e}^{\mathrm{i}\Big(\big(p_j^{\mathrm{(L)}}-p_j^{(0)}\big)\big(q-q_j^{\mathrm{(L)}}\big)+\Delta S_j^{\mathrm{(L)}}-\Delta S_j^{(0)}\Big)},
\end{eqnarray}
where $q_j^{(0)}$ and $p_j^{(0)}$ are respectively the positions and momenta of the field-free wave parcels, given by (\ref{wp_pos_mom}). As indicated by this transformation, the field-free wave parcel is first centered at position $q_j^{\mathrm{(L)}}$, evaluated from the classical trajectory. Its momentum and phase offset are then set in position space with the classically evaluated parameters $p_j^{\mathrm{(L)}}$ and $\Delta S_j^{\mathrm{(L)}}$, respectively. Thus, the transformation (\ref{classical_adjustment}) makes use of purely classical information to account for the control field. This classical information is sufficient to explain the effect of the control field on the fringes in the photoelectron spectra, as shown in Fig. \ref{clims}. Indeed, the spectra evaluated using our classical model represent a marked improvement to those erroneously predicted by the CVA (cf. Fig. \ref{cvaims}).% Our classical model thus provides valuable insight and predictive power.

For a given laser field strength, the fringe patterns are reproduced over a wide range of momenta, despite the fact that only a single initial momentum was used for each classical trajectory. The classical simulations neglect two purely quantum-mechanical effects: the influence of the control field on the reflection probability and the phase acquired upon reflection (i.e. the scattering phase shift for the backward direction). Consequently, the position and contrast of the spectral fringes predicted from our model is slightly off at larger control field strengths. These purely quantum mechanical effects cannot be explained by the classical model.

In order to clearly illustrate the key physics and to make the relevant effects more discernible, we considered a rather large system. Such a system might be a dissociating diatomic molecule, a dimer, an excimer, or a nano-structure composed of two spatially separated entities. Our analysis also applies to smaller systems. A smaller system would result in a broader spectral modulation, requiring a larger XUV bandwidth to capture enough fringes; or equivalently stated, it would require a shorter attosecond pulse so that the wave parcel is made up of two spatially distinct portions $w_\mathrm{D}(q)$ and $w_\mathrm{R}(q)$. Our approach applies more generally. For instance, in the case of a delocalized initial state, the starting points of the classical trajectories should be located near the peaks of the initial state, with all first-order re-scattering events considered for each trajectory.

%We have established a simple and intuitive model that can explain the main physics of laser-assisted IMS. If the attosecond pulse is sufficiently short to localize the outgoing electron wave packets, there is a transparent relation between the interference pattern in the photoelectron spectrum and the two trajectories taken by the outgoing electron. Thus the ultra-fast dynamics of a quantum system can be imaged straightforwardly, and with attosecond resolution, from the measurement of the photoelectron spectrum, which may constitute a novel spectroscopic method in attosecond science.

In conclusion, we have shown that an external NIR laser field controls the re-scattering of an electron, which can be observed by measuring the photoelectron spectrum for different NIR field intensities. The NIR field mainly affects three parameters of the re-scattered wave packet: it changes its momentum, its action and its apparent starting position, the latter of which corresponds to a \emph{delay} when considered in the time domain. On the other hand, for moderate intensities the control field hardly affects the scattering phase shift of the re-scattered electron.  Moreover, we found that the probability of re-scattering is affected by the strength of the control field. This might provide a means to generate and control a spatially and temporally confined electric current on a single atom by launching a free electron wave packet with an attosecond pulse in the presence of a controlled NIR wave form.

As evidenced in our study, the semi-classical Coulomb-Volkov approximation cannot describe these effects, and indeed breaks down for such a spatially extended system. In order to uphold a physically intuitive picture of laser-dressed scattering, we presented a new model based on two classical trajectories that quantitatively explains the influence of the NIR control field on the photoelectron interference pattern. Our model is generalizable to larger systems, and thus constitutes a powerful tool for interpreting this new kind of spectroscopic measurement, where a spatially extended system is monitored or characterized using its own outgoing electron.

%These parameters can be directly accessed, without the need for optimization, by Fourier-analyzing the measured NIR-dressed photoelectron spectra.

\begin{acknowledgments}
The authors are grateful to E. Goulielmakis, A. Kamarou, M. Korbman and N. Karpowicz for valuable discussions. This work was supported by the DFG Cluster of Excellence: Munich-Centre for Advanced Photonics.
\end{acknowledgments}

\appendix

%\bibliography{tau}% Produces the bibliography via BibTeX.
%merlin.mbs 2010-03-15 4.21a (PWD, AO, DPC)
%Control: key (0)
%Control: author (8) initials jnrlst
%Control: editor formatted (1) identically to author
%Control: production of article title (-1) disabled
%Control: page (0) single
%Control: year (1) truncated
%Control: production of eprint (0) enabled
%merlin.mbs 2010-03-15 4.21a (PWD, AO, DPC)
%Control: key (0)
%Control: author (8) initials jnrlst
%Control: editor formatted (1) identically to author
%Control: production of article title (-1) disabled
%Control: page (0) single
%Control: year (1) truncated
%Control: production of eprint (0) enabled
%

\end{document}